# Critical Review of Models, Containing Cultural Levels beyond the Organizational One

**Kiril Dimitrov**

**Summary**:

The current article traces back the scientific interest to cultural levels across the organization at the University of National and World Economy, and especially in the series of Economic Alternatives – an official scientific magazine, issued by this Institution. Further, a wider and critical review of international achievements in this field is performed, revealing diverse analysis perspectives with respect to cultural levels. Also, a useful model of exploring and teaching the cultural levels beyond the organization is proposed.

## 1. Introduction

Bulgaria's accession to the European Union from the 1[st] of January 2007 proved to be not only a glorious date, that marked general acknowledgements of the successful end in hard and tedious endeavors, exerted by local politicians, business leaders and society for establishment of an open market economy with democratic system, but also the beginning of an even more "complex game" within business and political contexts of the relations among separate member states in the Union. Just like the strained post-merger period of two previously separately existed companies, the elder member states, on one side (although there are many differences among them), and Bulgaria, on the other side, had to adapt culturally to each other, taking their time to surmount the typical cultural shock of becoming well acquainted with each other and long held prejudices of each other (Roth, 2012; Lewis, 2006; Paunov, 2009). Relying on the axiom "one who pays the bill, orders the music", it is not surprising that Bulgarian politicians, business leaders and society are doomed to walk the larger part of the road to "the point of cultural cohesion" (fit), since our developing country is a beneficiary of pre-accession funds and operative programs of European social funds for current and future program periods, and is subject to monitoring process by the European committee, measuring the results

*Kiril Dimitrov is a Ph.D., chief assistant at the Industrial Business Department of UNWE, e-mail: kscience@unwe.eu

of followed paces in required reforms in different spheres (business; ecology; legislation; judicial power; preventing nepotism and corruption practices, financial frauds, money laundering; security, electronic government, etc.). In fact the first five years of country's membership in the "Club of the richest" above all things represent a period when:

♦ Bulgarian business leaders and politicians had to redefine their concept of internal and external market, accepting the challenges of higher competitiveness and social responsibility of the entities in the Uniform European market. The last undertaking is embodied in necessary changes in the basic assumptions of the "local players", associated with redefining (updating) of their answers to sensitive issues, constituting the contents of the main problems in organizational culture, i.e. external adaptation and internal integration.

♦ The World financial and economic crisis had to be used as an opportunity of seamless entering into the regional markets of a number of other EU member states by Bulgarian business entities.

♦ Bulgarian business and politicians had to further develop their relations with the other EU member states, and especially these that are culturally closer to us.

♦ Bulgarian state had to accept the role of an outside boundary for the EU, diligently performing respective rights and responsibilities.

The achieved results on these criteria are far from being excellent, but undertaking interventions in these spheres, considering the accompanying cultural context, comes as the inevitable future of Bulgarian business, politicians and society, switching from unprofitable "nominal membership" in the EU to real, beneficial inclusion in the activities and deliberate development of the Uniform European market. That is why a greater attention is needed to studying of cultural levels beyond organization.

## 2. Dominating Views to Cultural Levels Across the Organization

Traditionally local scientists at UNWE do not put a great emphasis on knowledge and research in the field of cultural awareness and intelligence beyond the organization, excepting (adapted and partial) reproductions of Hofstede's survey of national cultures or Trompenaars and Hampden-Turner's cultural dimensions in Bulgarian context, done by Minkov (2002), Genov (2004), Davidkov (2005), Ivanov et. al. (2001), and Kolev et. al. (2009a). Outside the academic society, in the local business circles and public administration, pervasive negligence is demonstrated to cultural issues that may be considered with confidence as one of the main sources of the deep, long, local economic and political crisis from the last decade in the 20th century. A review of the articles, oriented to cultural studies, and published in the issues of the UNWE's scientific magazine, confirms this statement because there the research results revealed that the study of the cultural levels beyond organization has not been chosen as an investigative question by the scientists (see table 1).

*Table 1. A List of Articles in the Sphere of Organizational Culture, Published in "Economic Alternatives" Magazine (Previous Name – "Alternatives") at the UNWE[1].*

| Article's author and heading | Subject-matter, representing cultural levels (layers) |
|---|---|
| 1. Dimov (2010), Where Are We in the European Virtual Space? | - indirect relation to national level<br>- virtual culture |
| 2. Dimitrov (2009), Several Norms and Beliefs, Defining the Attitude to Human Resources in the Industrial Organizations | - indirect relation to national level |
| 3. Kolev, Rakadzhiyska (2009), A Tendency toward New Cultural Attitudes of Business Agents in Bulgaria | - national level<br>- the relation between national level and organizational one |
| 4. Andreeva (2008), The Cultural Industries in the Countries of Southeast Europe and their Economic Impact in the Context of Social Transformation | - indirect relation to national level and industrial one |
| 5. Paunova (2007), How to Characterize and Assess the Culture of an Organization | - organizational level |
| 6. Jankulov (2006), Surveying Organizational Culture of Trade Firms in Bulgaria | - organizational level<br>- industrial level |
| 7. Milkov (2006), Lecturer's Information Culture As a Factor, Creating High Quality Educational Service at Higher School | - indirect relation to professional culture |
| 8. Todorov (2006), Negotiation Strategies in Multicultural Business Environment | - indirect relation to national level and organizational one |
| 9. Alexandrova (2005), Entrepreneurial Orientation in the Context of National Cultural Environment | - indirect relation to national level<br>- entrepreneurial culture |
| 10. Dimitrov (2005), Conflictology and Conflictolocical Culture | - culture of the specialist |
| 11. Parusheva (2005), Destination Bulgaria in the Context of the Social and Cultural Effects of Eurointegration | - indirect relation to national level |
| 12. Spasov (2002), Institutional Change and Economic Transition | - indirect relation to national level |
| 13. Stavrev (2002), Bulgaria's Absurdities are Normal, But Only for Us | - indirect relation to national level |
| 14. Chankova (2001), Firm Culture – a Base for Efficient Innovation Activity | - organizational level |
| 15. Dimitrov (2000), Individualism of New Elites in Bulgaria and Development of Civil Society | - social strata<br>-national level |
| 16. Todorov (2000), The Relation Strategy – Structure – Organization Culture in Small and Medium Sized Firms | - organizational level<br>- basic assumptions are described through Schein's definition of organizational culture |
| 17. Peycheva (1999), Necessity of Ethics Defense in the Firm | - indirect relation to organizational level |
| 18. Zlatev (1997), Issues and Challenges, Confronting Industrial Managers in Bulgaria | - indirect relation to national level and professional culture |

[1] The articles are accessible either as hard copies in the university library (i.e. in the periodicals reading room) or as electronic documents through the university sub-site.

The contents of cultural awareness courses at UNWE traditionally present the points of view of one or more of the following scientists: Hofstede, Paunov, Trompanaars and Hampden-Turner, and Todorov. Hofstede (2010a, 2010b) justifies the existence of three levels of culture, influencing to a great extent the formation and evolution of a target organizational culture, as follows: national level, professional level, and gender level (see figure 1).

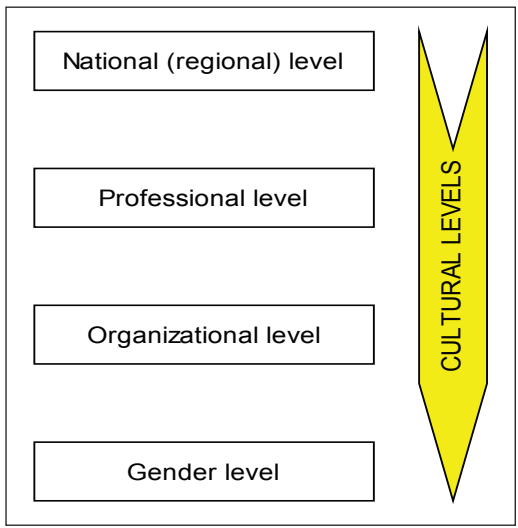

*Fig.1. Hofsede's View of Cultural Levels beyond the Organization.*          *Source: Hofstede (2010a)*

In his framework of cultural levels Hofstede establishes a hierarchical (vertical) order, locating at the highest point the attributes of the national level, followed consecutively by professional, organizational and gender one. According to the scientist national cultures differ from each other on the basis of unconsciously held values, shared by the majority of the respective population. He defines a value as a widely preferred state of being in comparison to other possible ones. In this way the Dutchman proposes acceptable explanations to the observed cultural specifics at national level, as follows:

♦ National cultures show their stability in time, permitting changes in dominating value set not until occurrence of an evident shift among generations.

♦ The turbulent influences of the environment may cause in most of the cases just changes in the practices (symbols, heroes, rituals), while the underlying values still remain intact. The last ones may undergo certain changes in extreme occasions, such as wars, death of a family member, severe illness, natural disasters, etc.

♦ The cultural unit at national and regional level may not correspond to the established boundaries among different countries. So, not only culturally similar regions may belong to different states, but also big countries may consist of culturally differing regions. The smaller the country, the greater cultural homogeneity it possesses.

The importance of the professional level stems from individual's obligatory "mind programming" before his/her entering in a certain occupational field. Hofstede views it as a mix of national and organizational cultural elements that determine its place in the proposed hierarchy. Occupational cultures possess their sets of specific symbols, heroes and rituals.

Hofstede compares the intensity of expressed feelings and fears in regard to demonstrated behaviors by the opposite sex to the intensity of people's reactions to clashes with foreign (alien) cultures. He assumes that national culture influences to a great extent the established difference between sexes. The existence of two differing male and female cultures in a given society provides a plausible explanation of the reason why the traditional division of gender roles is not amenable to deliberate changes. Both males and females may possess required knowledge, skills and capabilities to perform a certain job, but the representatives of one of the sexes may not be in congruence with the traditional symbols, may not resemble familiar heroes, may not share established rituals, or even may not be adopted in this different role by the opposite sex in the organizations.

Hofstede et. al. (2010b) enrich a bit the list and descriptions of proposed cultural levels, as follows:

♦ The national level represents the aggregate of people, living in their mother country and the immigrants. The culture of the representatives of the last group may be considered as a specific mix of national cultures according to the countries these people migrated during their lifetime.

♦ The regional level is further segmented to reflect diversity in ethnic and/or religious and/or linguistic affiliation, since the majority of nations consist in motley groups.

♦ The generation level appears to explain potential differences among grandparents, parents and children.

♦ The social class level is introduced to assess potential differences, related to available educational opportunities and to a person's occupation or profession.

In his turn Paunov (2005) explores the relation of organizational culture with other cultural systems, differentiating culture (specific to group or category and learned) from related terms as human nature (universal and inherited) and personality (specific to individual, inherited and learned), adhering to Hofstede's vertical perspective (1994). But Paunov's interpretation reveals certain nuances (see figure 2). *First*, the hierarchical perspective in studying and organizing of cultural levels is preserved and a bit enriched, concerning the position, determined for the individual who is viewed dually as a member of civil society and as a performing employee, contributing to a certain organization. *Second*, the direct link between "individual culture" (as a rule considered as the basic cultural unit, not a cultural level) and "organizational culture" implies author's support to the existence of an "overarching organization culture", consisting in some cases of separate sub-cultures – a notion, shared by the majority of researchers in the cultural field. Additionally, this approach implies Paunov's modern concept of career as individual's portfolio of skills and competences that may be transferred from one workplace to another (Arthur, Rousseau, 1996). It may be concluded that the author inherently considers the possibility of hired laborers' outliving employer organizations. *Third*, a dashed arrow is constructed to show Paunov's adhering to Hofstede's vertical perspective, revealing the subordination of "organizational culture" to "national culture". *Fourth*, Paunov labels all identified cultural levels beyond the organizational one with the collective term "other cultural systems". *Fifth*, diverse criteria for further segmentation of the elements, belonging to the identified main cultural levels (in the text boxes), are proposed (the right side of the figure). The greater importance of some of these criteria is underlined by their use in more than one cultural system (i.e. a level beyond the organization). *Sixth*, the "industrial segmentation criterion" is located in the upper right from the organizational culture level (text box) to reveal the logic direction of subordination.

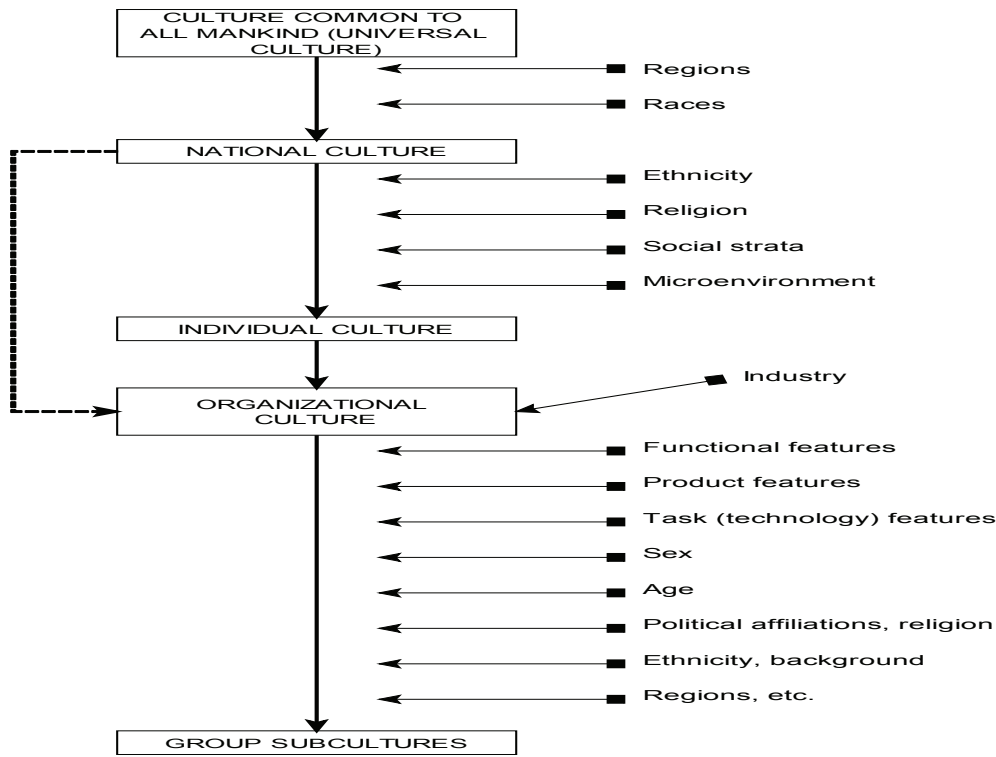

*Fig.2. The Relation of Organizational Culture with Other Cultural Systems*

Additionally, Paunov (2008) enriches his perspective on cultural levels beyond the organization in order to analyze in a better way business-related cultural issues by presenting Trompenaars and Hampden-Turner's framework of cultural layers (1998). The notion of hidden from all sides cultural layers is directly illustrated here (see figure 3). But again here the analysis inherently goes on national level (cross-cultural business communications, business culture), although it permits making the logical conclusion of the existence of a strong relationship between national and organizational levels. In fact Trompenaars and Hampden-Turner (2008) use in the examples, accompanying the explanation of the proposed framework, stories of people's experiences while entering different countries and communicating with different ethnical groups (for example Burundi, Hutus, Tutsis, Japanese, Eastern Europe, Dutch, Central Americans, etc.). Traditionally for the applied by them "union approach" of presenting cultural layers, the scientists start their analysis in the direction from outside to inside, because they consider that concrete factors constitute one's first experience of an alien culture, i.e. the level of explicit culture. The last one is filled up by certain contents, as follows: "the observable reality of the language, food, buildings, houses, monuments, agriculture, shrines, markets, fashions and art". All these cultural elements are grouped as symbols of deeper cultural levels. The great inaccuracy of expressed opinions and made conclusions about a given culture, based just on these items, is heavily underlined by revealing the impact of individual's prejudices, reflecting predominantly his/her background, not the assessed community.

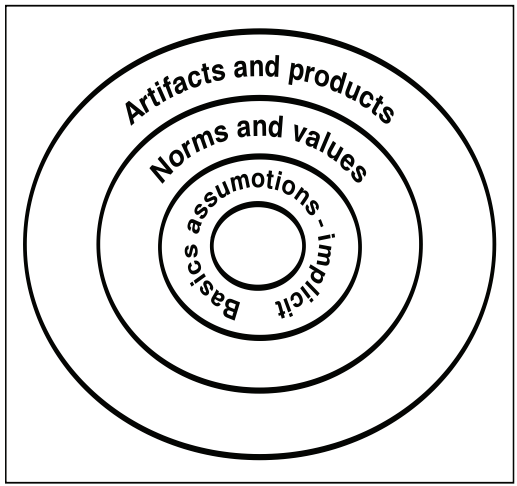

*Source: Trompenaars, Hampden-Turner (1998),*
*Paunov (2008)*
*Fig.3. A Framework of Cultural Layers*

Unbiased observer's posing direct questions to members of the assessed group, in order to decipher strange and confusing behaviors for him/herself, is considered as a normal way in penetrating into a deeper cultural layer, i.e. the layer of norms and values. Here the authors describe traditional definitions and characteristics of norms (what is right or wrong, "how I normally should behave") and values (what is good or bad, an attribute close to shared ideals by group members, "a criterion to determine a choice from existing alternatives") and a classification criterion of one of these terms (formality: formal norms – written laws; informal laws – social control). The congruence between norms and values is greatly appreciated by Trompenaars and Hampden-Turner who see it as the main source of cultural stability, the antipode to which leads to "a destabilizing tension" and "disintegration". The stability and salience in shared meanings of norms and values in a community are considered as a sufficient condition for their further development and elaboration. Here, it is assumed that dominating norms and values influence group members on both conscious and subconscious level.

The core layer of assumptions about existence is used to explain the great diversity among different groups of people in preferred definitions of norms and values, initiating the analysis with defining of survival as people's most basic value, in terms of "fighting" with nature on a daily basis and giving examples again with the specific problems of different nations and taking into account the specific conditions of different geographic regions and the available resources to the inhabitants. Trompenaars and Hampden-Turner rely on the axiom that people tend to self-organize in order to maximize the effectiveness of their problem-solving processes. It seems these problems are complex and recurring (revolving) and require continuous efforts in their daily resolution. The repetitive character of these actions, undertaken by people, for the sake of achieving short term successes in a certain field, causes their gradual disappearance from human consciousness. In this way individuals find their ways of coping with anxiety by reacting to environment threats with proven solutions out of their awareness, i.e. the set of shared basic assumptions by the group. Discussion is determined to be the only one technique that may bring to surface a basic assumption by posing the right questions which provoke as a rule confusion or irritation among target participants. Then surprisingly the authors claim that this is the way how organizations work, explaining even the reason of initiating change programs in the entities, i.e. situations when "certain old ways of doing things do not work any more".

The earliest book in the sphere of firm culture at UNWE, edited by Todorov (1992), is among the most popular sources, cited by students in the course works, assigned to them through the enacted syllabus, but its contents does not provide detailed information of cultural levels outside the organization (see table 2). So, students have to put in additional research efforts, searching for related sources in the scientific databases.

existing relations, application spheres, assumptions of layer contents, orientation to (re)(ab)solving certain issues, etc. The hierarchical perspective in structuring cultural levels is further elaborated to include the global one (Erez, Gati, 2004; Wilhelms, Shaki, Hsiao, 2009). The option of lateral relations among levels is also presented, through the lens of Karahanna et.al. (2005) who even defines a supranational level as a mild equivalent of global one. Hofstede's work on cultural levels is revealed not only through the criterion of their scope or generality, but also by the essence of relation between values and practices, expressed by their different mixes (Hofstede, 1990; Hofstede, 1991; Hofstede, 2010b). The multiple attitude to individual is described, too, outlining three alternatives: (a) as just a product of a certain culture (Wilhelms, Shaki, Hsiao, 2009) or (b) as a separate cultural level (Erez, Gati, 2004), or (c) as a set of two layers (Espinar, 2010). Specific sets of cultural levels,

*Table 2. Todorov's Penetration into the Firm Culture Sphere.*

| Book | Subject-matter, representing cultural levels (layers) |
|---|---|
| Todorov, K., (editor) Firm culture and Firm Behavior, Publishing House "VEK22", 1992. | - (pp. 22-42) **Hofstede's** levels: national culture, organizational culture (symbols, heroes, rituals, values), regions, social strata, professional culture (the last two items are not called cultural levels);<br>- (pp. 43-59) **Henze's**: (a) microculture (firm culture and subcultures); (b) macroculture (national, international and regional); (c) Schein's framework of organizational culture (artifacts, values, assumptions);<br>- (pp.84-115) **Kleinberg's:** social culture, cross-cultural level and international culture; |

## 3. The Broader Horizon of Frameworks, Describing the Cultural Levels beyond the Organization

Scientific databases (EBSCO, ScienceDirect, Springer, Scopus and ProQuest) provide a richer picture on cultural levels as diverse perspectives of analysis, essence of depending on different application spheres as higher education (Rutherford, Kerr, 2008), information technologies (Ali, Brooks, 2009), task forces (Schein, 2010), human learned behavior (O'Neil, 2006) are analyzed. Even perfunctory attention to cultural levels concept is detected (Schein, 2010).

*3.1. Erez and Gati's Model of Cultural Layers*

Erez and Gati (2004) propose a five level cultural model, bearing its specific structural and dynamic dimensions (see figure 4). The model is structured as a hierarchy of layers, consecutively nested one in the other. The core level in the framework is represented by the cultural image of the individual who participates in groups, organizations, nations

♦ A society's shared meanings are accepted to a satisfactory extent by the running top-down socialization process. In this way societal values become a part of an individual's personality.

♦ The components of higher rank systems (group, organization, national level) are constructed by value aggregation and sharing process.

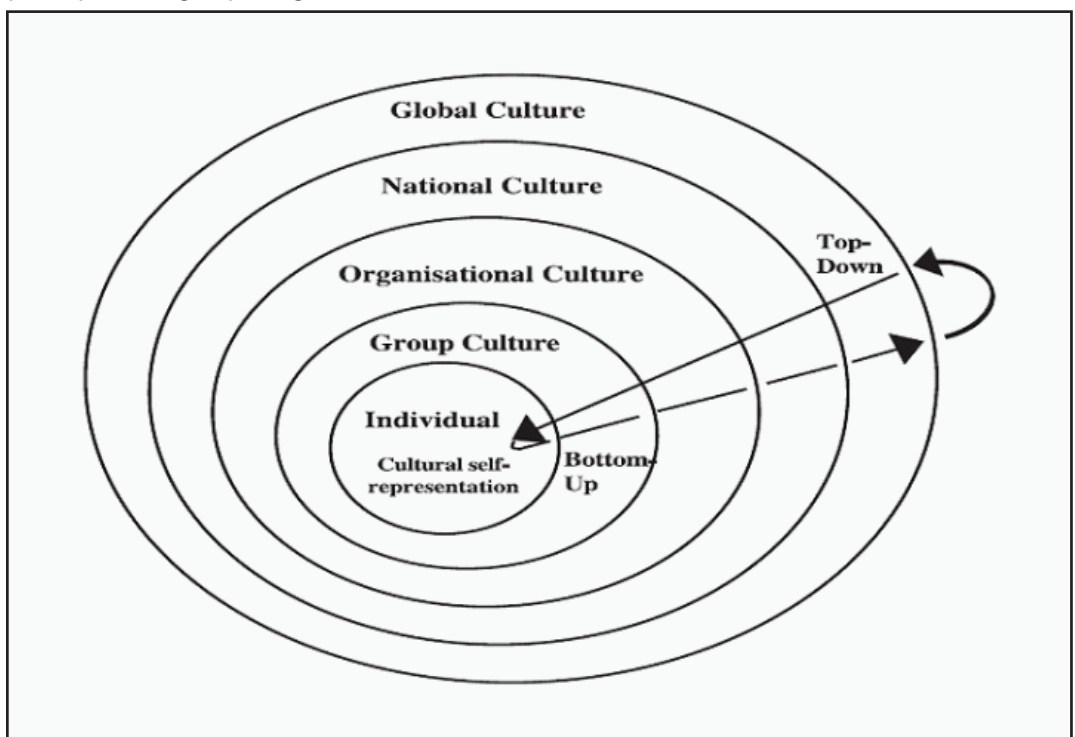

Source: Erez and Gati (2004)

*Fig.4. Erez and Gati's Model of Cultural Layers*

and global culture. The authors view at culture as a system of shared meanings that may emerge at each of the mentioned levels. Model's dynamics is achieved by maintained relations among the levels, especially the ways in which they influence each other. The last may be summarized, as follows:

Thus, Erez and Gati succeed in defining important cultural characteristics of the mentioned cultural levels:

♦ *Global culture.* The contemporary working environment is totally transformed by occurred globalization, expressed mostly by the increasing economic inter-dependency among countries, caused by international flaws of commodities, services, capital, technologies, and people. The "western culture" established on this level dominating core values as: freedom of choice, free markets, individualism, innovations, tolerance to changes, diversity and interconnectedness. Globalization impacts the other cultural levels by means of top-down processes. Vice versa, the bottom-up processes may bring to the surface of global culture a new characteristic that in fact embodies shared behaviors and norms by the members of the lower rank cultural levels. In this way the homogeneity of global culture level increases. Exposing individuals to global culture influence creates their global identity. The acceptable fit between global and local identity is a prerequisite for successful adaptation to both environments. That is why the scientists claim that members of national cultures, characterized by high individualism, low power distance, and low uncertainty-avoidance may better adapt to the global work environment in comparison to people with other cultural background.

♦ *National culture*. A nation and/or a state represent the analyzed cultural unit at this level. Differences among the chosen units are sought with respect to held or professed national values or organizational practices and behaviors, event management, effective leadership characteristics, basic axioms, and some social behavior theories. Additionally, the achievement of consensus on desired and dominating values in a given society at least partially establishes the boundaries of the unit at national level.

♦ *Organizational culture*. It is defined as the shared set of beliefs and values by the members of a certain organization, influencing demonstrated behaviors. The scientists propose the following basic cultural dimensions at this level: orientation to innovations, attention to details, orientation to achieving outcomes, attitude to risk-taking, and focus on teamwork. Additionally, it is accepted that homogeneity of personnel members' perceptions and beliefs to a great extent determine the strength of a certain organizational culture.

♦ *Group culture*. It is reflected by shared values by the members of a certain group. Orientation to shared learning, psychological safety in the team when expressing one's own doubts, interpersonal trust and support are the attributes, constituting the set of the most important values at this level.

♦ *Individual level*. The scientists consider at this level the own cultural performance (disclosure) of the person, structuring it as a mix of collectivistic values and individualistic values, incarnated in the self.

### 3.2. Wilhelms, Shaki and Hsiao's Perspective on Cultural Levels

The standardized model by Wilhelms, Shaki and Hsiao (2009) for classifying culture reveals its multidimensional dynamics. By means of extensive review of accessible scientific literature and its successive logic cataloguing the scientists outline the existence of five cultural layers, as follows: micro culture, meso culture, macro culture, meta culture, and global culture (see table 3).

*Table 3. Cataloguing of Scientific Literature, Aiming Subsequent Identification of Cultural Layers*

| CULTURAL LAYER | DIFFERENT AUTHORS' CONTRIBUTIONS TO THE EMERGING CONTENTS OF CERTAIN LAYERS WHICH ARE ORIENTED TO: |
| --- | --- |
| Micro culture | 1. Study of health care, especially dealing with cultural influences in providing such care.<br>2. The sphere of organizational culture.<br>3. Exploring the role of firm constituent interactions.<br>4. Rethinking Caribbean families in the dimension of extending the links.<br>5. Study of immigrant adolescents who behave as culture brokers.<br>6. Study of strategic divestments in family firms, especially the role of family structure and community culture.<br>7. Measuring the impact of values in a concrete company. |
| Meso culture | 1) Study of professional and industry culture.<br>2) Exploring of the micro, meso and macro terms in marketing context.<br>3) Study of the pitfalls of family resemblance, especially investigating the reasons why transferring planning institutions between similar countries is delicate business.<br>4) Study of personal and political agendas, pursued by women managers in Hong Kong.<br>5) Study of subcultures of consumption, especially an ethnography of the new bikers.<br>6) Exploring the engineer's perspective in marketing.<br>7) Exploring clothing stories: consumption identity, and desire in depression-era Toronto. |
| Macro culture | 1) Study of national framework, defined by geographic boundaries.<br>2) Exploring socio-cultural factors that influence human resource development practices in Lebanon.<br>3) Study of extending the cultural research infrastructure, especially the rise of the regional cultural consortiums in England.<br>4) Exploring the role of national culture in international marketing research.<br>5) Study of market orientation and the property development business in Singapore.<br>6) Exploring the impacts of some organizational factors on corporate entrepreneurship and business performance in the Turkish automotive industry. |
| Meta culture | 1) Study of culture in multinational organizations or the so called cross-cultural patterns.<br>2) Study of international differences in work-related values.<br>3) Exploring the levels of organizational trust in individualist versus collectivist societies.<br>4) Exploring the cross-cultural perspective on artists' attitudes to marketing.<br>5) Study of institutional panethnicity, especially boundary formation in Asian-American organizations.<br>6) Exploring the values in the West within the theoretical and empirical challenge to the individualism-collectivism cultural dimension. |
| Global culture | 1) Study of the biggest and the newest cultural layer.<br>2) Identifying global and culture-specific dimensions of humor in advertising.<br>3) Exploring the role of global consumer culture, regarding brand positioning through advertising in Asia, North America, and Europe.<br>4) Exploring young consumers' perceptions of multinational firms and their acculturation channels towards western products in transition economies.<br>5) Study of international organizations, the "education-economic growth" black box, and the development of world education culture.<br>6) Exploring the effects of culture and socioeconomics on the performance of global brand image strategies. |

*Source: Wilhelms, Shaki and Hsiao (2009)*

The scientists use the term "cultural layer" to describe the environment in which organizations perform. The individual – the smallest structural, cultural element (unit) – is located at the core of the model as a basic participant in the realization of all cultural layers. By contrast with Erez and Gati (2004), Wilhelms, Shaki and Hsiao (2009) deny the existence of specific individual culture, considering that the individual may only belong to (or possess) a certain culture, spread at least among the members of a certain group. So, it becomes evident that culture cannot be classified at "individual level". The proposed cultural layers embody possible locations of the organizations up and down the established environments in the model (see figure 5). The content of the separate layers is constantly changing. The information may be transmitted across the layers in two directions – from the surface to the individual and from the core to the global layer. Inside-out transformation occurs in the model when its core is affected (influenced) by transmitted information. All cultural layers are characterized by their density and erosion. Each cultural layer possesses specific density, corresponding to the achieved extent of structural complexity, incarnated in the present social and organizational forms in it. The density of a certain cultural layer is measured both by the penetration speed the information flows in it and by the ways in which information is being filtered during its flow. Cultural layer's density may change under the influence of external and internal factors (for example enacted laws and regulations). The availability of high density in a cultural layer means that information cannot penetrate in it because of enacted policies, normative acts, etc. In this way the two-way informational flow between the layers may be ceased with a set purpose – the flow's dynamics is blocked, its influence is decreased and consequences, associated with transmitted information, are prevented from occurrence.

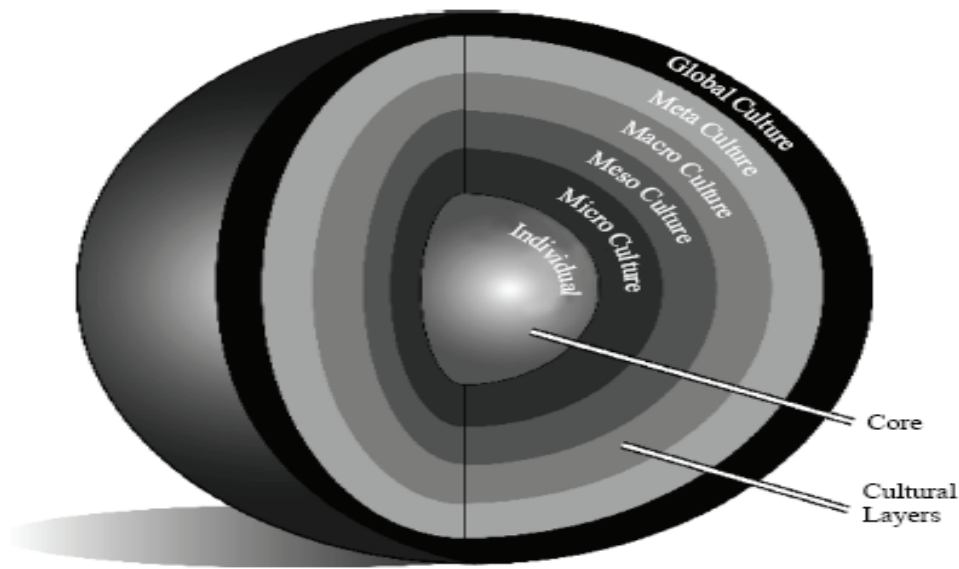

*Figure 5. Wilhelms, Shaki and Hsiao's Framework of Cultural Layers*

The cultural erosion is an intrinsic, naturally process, running in the model. Generally it is perceived as some kind of change. The erosion may come into being at each cultural layer as a result of influences from within it, due to the passing inside-out or outside-in informational flow, enacted laws and policies and natural changes.

Subculture is another basic attribute in this model and is identified as a certain social or organizational group that belongs to a given cultural layer which may be inhabited by many subcultures. Wilhelms, Shaki and Hsiao (2009) allow comparing different subcultures only if the last exist in a single layer and there is an acceptable fit between the explored cultural layer and its subcultures.

Thus, Wilhelms, Shaki and Hsiao (2009) become capable of precisely determining the essence of the core and each of the identified cultural layers, as follows:

♦ *The core, i.e. individual.* Individuals are products of culture they belong to. They have learned to act in specific ways within the socio-cultural environment. That is why they occupy the core of culture with relation to shared beliefs, attitudes, norms, roles and values. In this way the shared elements of subjective culture emerge and last, being transferred between generations as memories of personal experience including language, time, and space (layout). Therefore, the relation between memories and culture may be determined to the effect that the culture to a given society is "the same as" memories to an individual.

♦ *Micro culture.* This layer emerges by the formation of small groups (for instance friends, followers, etc.), although great volatility in group's size is possible, limited in the interval from the number of members in a family to personnel's average annual number in a company. Subcultures within the organizational culture context are accepted as "values" in this interval, too. Each group, which members hold shared behavioral patterns, may be a part of this cultural layer including ethnical groups. Single representatives of micro culture are components of wider cultural constructs and higher rank cultural systems. At this level the strongest connection (intimacy) among the members of a group may be achieved. The option of cultural segmentation is available here.

♦ *Meso culture.* It is intended to fill up the vacuum between micro and macro cultural layers. The authors share the concept that for sure meso culture differs from subculture. That is why they define precisely cultural groups, typical for the meso layer as larger than the average number of the personnel in a company and smaller in number than a nation, for example communities, consisting of two or more firms (consortiums, companies with at least two separate business units), or consisting of two or more families (entrepreneurial networks). The group members of meso culture are characterized by homogeneity, expressed in some behavioral aspects.

♦ *Macro culture.* The scientists define it as collective programming of human mind in a specific geographical region with upper limits, coinciding with territories of separate states, i.e. the national framework. Thus they succeed in restricting spheres and dimensions of undertaken studies at this layer. The macro culture systems take the shape of clusters from organizations or social groups within a target national framework, for example two or more representatives of meso layer which may not share the same attitudes but obligatory belong to a given national framework. That is why the authors conclude that the achieved group homogeneity here is weaker in comparison to meta culture layer.

♦ *Meta culture.* Its structure comprises of two or more national cultures including their social groups, or multinational geographical region or alliance (for example: European Union, North American Free Trade Agreement, etc.). Meta cultures are characterized by existing behavioral similarities, discovered in two or more macro (national) cultures. Meta cultures are also influenced by mighty organizational alliances, functioning within them. The authors provide the example of regional industries that are presented by two or more national industry sectors, operating in two or more countries.

♦ *Global culture.* The whole planet is the magnitude of the construct that is widely used by media, academics and business, marking the made cultural choice on worldwide scale. This is the external border of the model. Two or more continents represent the smallest inhabitant in this cultural layer. Large organizations, operating within multiple national boundaries on the terri-tory of two or more continents are labeled as global. That is why the scientists claim that the shared beliefs patterns by global organizations compose the contents of global culture. The scientists use an appropriate example to illustrate manifestations of global culture as functions, performed by the World Trade Organization in connection with development of policies for business organizations on a global scale. The World Bank, The International Monetary Fund, The United Nations Organization may be accepted as other entities, influencing on a global scale.

Wilhelms, Shaki and Hsiao (2009) guarantee that their framework is not static by considering organizations that may be classified in a given layer at a definite moment as dynamically developing systems whose growth strivings may lead them to necessary penetration into other layers. So it sounds logical that localizations (spheres) within which a given company or industry operates, and the achieved phase of their evolutionary development are the two criteria that predetermine their belonging to a certain cultural layer. That is why the scientists ground the design and implementation of a regular monitoring review and assessment process, concerning the current performance state of each target organization at a given cultural layer.

### 3.3. Other Necessary Facets in Hofstede's Work

The results from two large scientific projects, led by Hofstede, help the scientist identify and organize in a sequential order different cultural levels, based on the crite-rion of respective weight in "the ratio" between values and practices (symbols, rituals, heroes). Thus, the Dutch reaches the conclusions that national cultures differ to the greatest extent in values, dominating in each of them (Hofstede, 1991; Hofstede et. al., 2010), while organizational cultures differ to the greatest extent in practices, dominating in each of them (Hofstede et. al., 1990). In this way it becomes clear that different mixes of values and practices exist at the identified cultural levels beyond the organizational one (see figure 6).

undergone as an adult (above 18 years, for professionals with higher education at the age of above 22 years) whose individual set of values is already formed. *Third*, gender and nationality are the only two cultural attributes, present at birth in one's lifetime. *Forth*, values of founders and leaders in a given organization may differentiate from the individual value sets of the employees. That is why founders and leaders may create and disseminate only daily practices among the organization's members, embodied in symbols, heroes and rituals. Since the em-

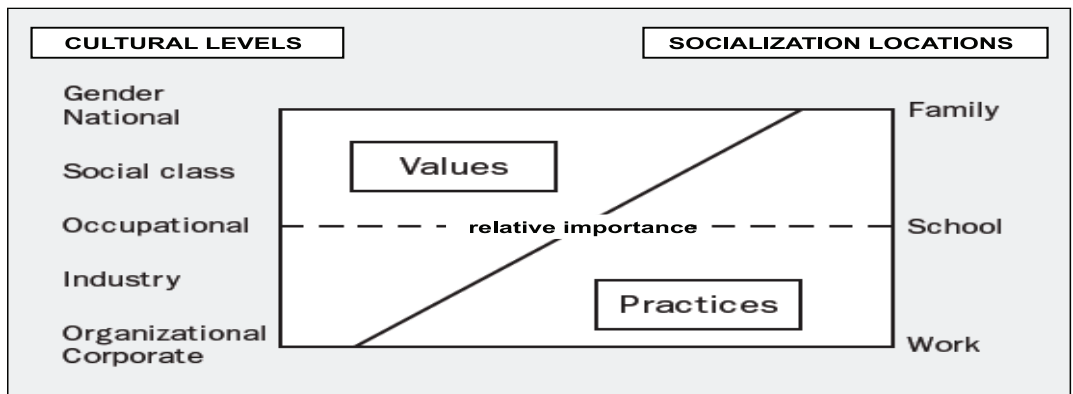

*Source: Hofstede et. Al. (2010)*

*Fig.6. Hofstede's Cultural Levels and Corresponding Socialization Locations*

Hofstede explains the availability of differences in "values – practices" observed mixes at separate cultural levels with diverse locations of learning or socialization of the mentioned couple of attributes. *First*, values as "mental programs" are acquired mainly during childhood and adolescence in one's lifetime and in locations as one's family, neighborhood and school. *Second*, the organizational culture is learned by a newcomer in a certain entity during the process of his/her socialization which is generally ployees have their personal and social lives out of the organizations, they are not stimulated or forced to change the items in their established individual value sets in contrast to the inhabitants of a prison, a mental hospital, an orphanage, a monastery, a nursing home, etc., where the members are isolated from the wider community for a certain time, reside and/or work on a confined territory and generally are in a similar life situation, leading a formally administered way of life. It seems evident that membership in a given organization is not the prime criterion of formation and/or changes in an individual's values (see figure 7).

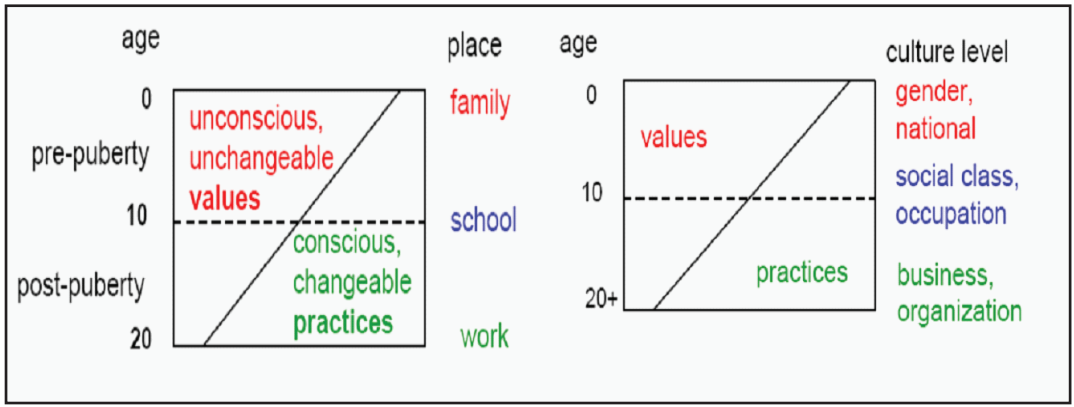

*Fig.7. Hofstede's Concept of Forming Values in One's Lifetime*

That is why the specialists in the sphere of human resource management are considered as the main driving force in:

♦ Maintaining or changing the dominating values in the organization by recruitment and selection of the "right people" for the organization or laying off "inappropriate people" (gender, nationality, social class, education, age) and

♦ Ensuring the adequate adoption of the practices (symbols, heroes, rituals) by the newcomers through an established formal socialization process.

Therefore Hofstede infers that because of practices' domination over values in the mix, the organizational culture seems to a greater extent manageable through management's undertaking changes in the sphere of practices, i.e. preferred ways of employees' doing things in the organization, including facts about the business, how it works, proven cause-effect relationships, etc. The Dutch points out that as a rule an employer is not able to provoke changes in employee's values, since the last are formed earlier before their encounter, but may only activate already possessed by the subordinate latent values by allowing manifestation of previously forbidden practices in the organization.

### 3.4. Karahanna et. al's Special View of Cultural Levels

Karahanna, Evaristo and Srite (2005) assume that specific individual's characteristics account for potentially different influences that the separate cultural levels exert on him. Thus, the scientists succeed in identifying two streams of cultural level influence on individual's behaviors:

♦ The national cultural level has the mightiest impact when demonstrated behaviors are socially oriented or dominated by terminal or moral values.

♦ The organizational and professional levels have the mightiest impact when demonstrated behaviors are task-oriented or implicate competence values or practices.

In this way the scientists support the view of Hofstede (1991) and Straub et. al. (2002) that there exists a kind of interaction among the cultural levels, since manager and employee behaviors may be affected by diverse influences of different cultural levels (national, organizational, subculture, etc.). But Karahanna, Evaristo and Srite (2005) not only think that cultural levels and cultural layers[2] related to each other. Thus, the researchers reject previously dominated Hofstede's concept of hierarchical understanding as the only one perspective in analyzing and classifying cultural levels on generality criterion within the interval, formed by the upper limit of a (supra)national level[3] and the lower limit of a group one (see figure 8).

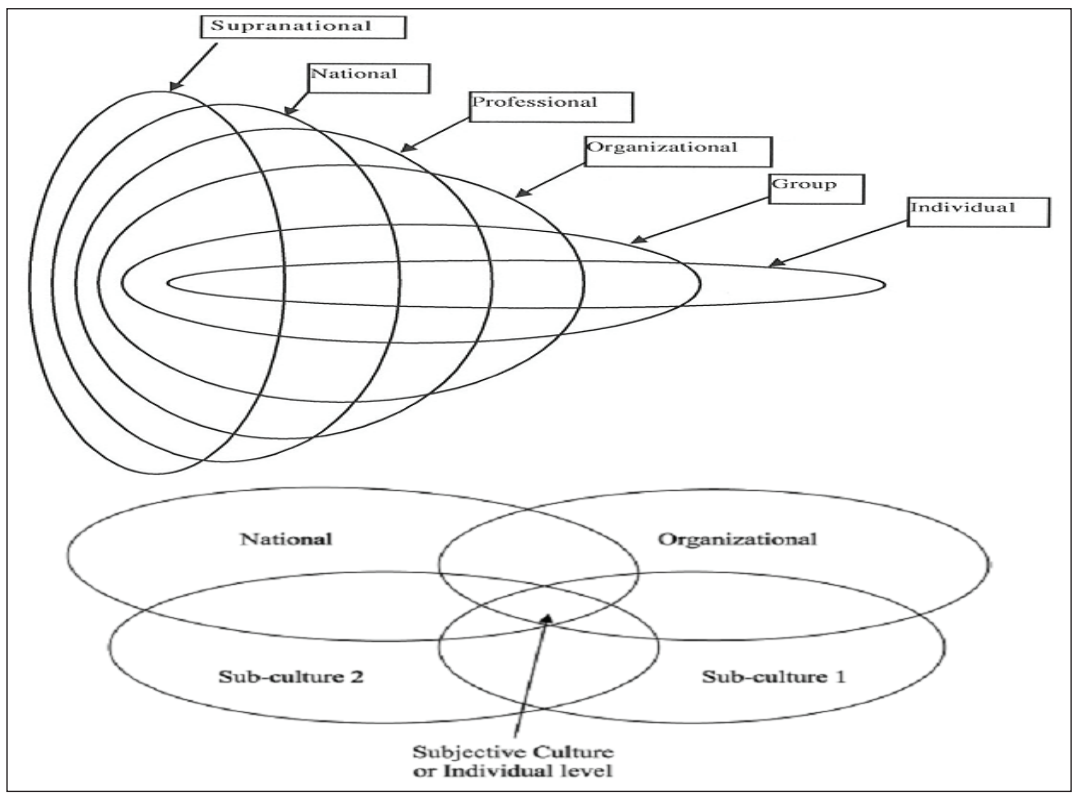

<div align="right">Source: Karahanna, Evaristo, Srite (2005)</div>

*Fig.8. Interrelated Levels of Culture – a Lateral Perspective*

may impact each person differently, based on particular situation and individual's values, but also emphasize the complex essence of this impact by bringing forth the idea that cultural levels to some extent may be laterally

The logic of scientists' presumption is supported by the essence of the contemporary business relations, as follows:

♦ The greater part of the multinational companies operate under specific condi-tions in which the overarching organizational culture encompasses a number of national, professional and other (sub)cultures.

[2] Hofstede's et. al. (1990) describe as cultural layers: symbols, heroes, rituals and values.

[3] Karahanna, Evaristo and Srite (2005) constitute the supranational level to describe any kind of cultural differences that may cross national boundaries or may be detected to exist in more than one nation, originating from regional, ethnic, religious and linguistic peculiarities.

♦ The complexity of collaboration forms among business entities predetermines the increasing number of cases when groups (task-forces) are created whose members: (a) may be permanently employed in different organizations; (b) may have different professions; (c) may derive from different nations and/or ethnicity; and (d) may profess different religions.

Since each individual is characterized by certain national, ethnical and language affiliations, religious orientation, special education, etc. which allow his being classified in different subcultures of the society, Karahanna, Evaristo and Srite (2005) may define the subjective culture of each person as a compound of at least several cultural levels. Additionally, if the individual works in a given organization, this entity is also presented in society by its dominating culture (see figure 8).

Karahanna, Evaristo and Srite (2005) share Hofstede's concept of two types of human learning in one's life – unconscious one, focused on values, and conscious one, focused on practices, but they go further to apply the system approach in analyzing values, considering their interdependence, the potential existence of a relative priority for each item in the set and assuming its stability. They leave an open door to changes in value set over time, due to migration, personal experience and extreme circumstances. Change management perspective to organizational analysis leads the scientists to conclusion that change agents find it easier to transform practices in comparison to values which is described as hardly feasible undertaking. In this way they pose the issue of property (essence) and quality in "value-practices" relation and outline some of its important shades, as follows:

♦ Practices make an impact on values during the formation period of the last, i.e. human childhood and adolescence.

♦ The practices entirely lose their influence on values during the later stages in human life.

♦ As a source of potential cultural differences values show greater relative importance at the more general cultural levels (for example the national one), while practices dominate at less general cultural levels (for example the group one).

♦ Practices are constantly changing because these attributes are to a greater extent related to the environment.

♦ Numerous interruptions in high degree "values-practices" fit (congruence) are observed. In fact in many real life cases the practices simply do not embody the underlying values or are not in congruence with them, which may be due to a great contradiction between the assigned practices at a given cultural level (for instance organizational one) and the values, filling with contents another cultural level (for instance national one).

### 3.5. Rutherford and Kerr's Framework of Cultural Levels

Rutherford and Kerr (2008) also propose a model of some kind of laterally interacting cultural layers that exist under the specific conditions of the respective internet culture in a given educational organization (see figure 9). They consider that the successful design of online training environment requires taking into account the specificity of certain teaching-related cultural layers as follows: a national culture, a professional-academic culture, dominating culture among the students in a given educational institution, a target classroom culture, the organizational culture of the analyzed educational institution and cultures, related to international education. By this model the scientists pay great attention to important cultural aspects in design process of online learning environments, i.e. educative software and online training computer plat-forms. Generally, design process of such deliverables flows within a certain cultural context whose values and standards get embodied in the deliverables. Authors' view lies on research results, revealing that deliberate increase in working speed, openness in communications, levels of informality in realized relations and freedom of questioning dominate in design of communications platforms in the region of North-America (Reeder et. al., 2004). But Rutherford and Kerr (2008) assume that the expression of such values may turn out inacceptable within certain cultures. That is why they consider the demonstration of intercultural competence as extremely important and outline it as developing a certain attitude to curiosity and openness, accumulating knowledge of the interactions among people in a society and acquiring skills for critical interpretation of new cultural knowledge. That is why the authors claim that the creation of online training environment, recognizing the influence of dominating cultural dimensions on instructors' and trainees' perceptions proves to be a very difficult task. The scientists try to resolve this issue by applying Tylee' set of subsequent design process decisions (2002) as a necessary condition in the design process of online training environment, concerning the acceptable extent of interaction among individuals, the appropriate motivational approaches for users, the establishment of a healthy balance between group and individual opinions or one-person work and teamwork, assessment of appropriate extent of uncertainty avoidance, choosing an appropriate teaching style for a target audience, selection of appropriate appraisal methods, etc.

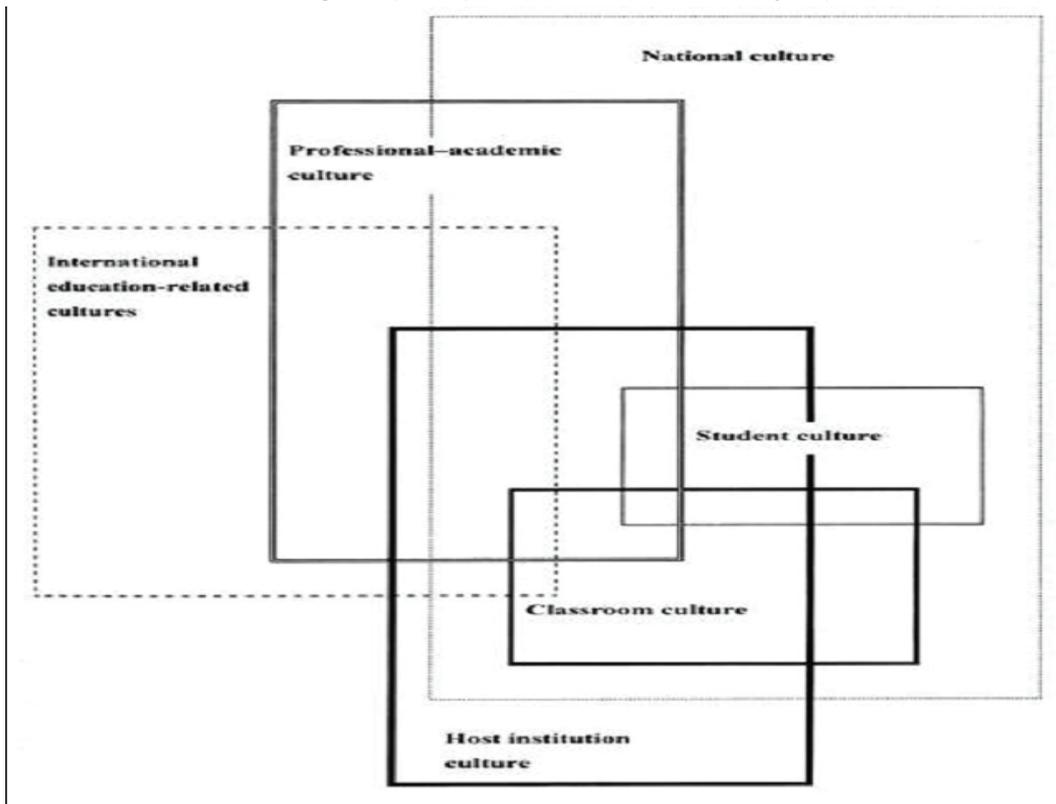

*Fig.9. Important Cultural Layers in Process of Online Learning Environment Design*

### 3.6. Espinar's Cultural Levels

Espinar (2010) applies a framework of cultural layers through the lens of international business communications, justifying this approach with already proven dependence of a company's successful performance on the global market on efficient and effective demonstration of communication ability in an intercultural context. The scientist considers that individuals transfer their cultural values to the communicative process. In this way she is able to explain the specific process of message filtering through a set of cultural layers, done by each (potential) business partner in cross-cultural business related encounters. Further on the researcher points out that demonstrated complexity of human behavioural modes determines potential success of undertaken business initiatives. In fact the applied framework of cultural layers, affecting the communication process, is summarized by Targowski and Metwalli (2003) who arrange the components in a sequential order by diminishing generality, define the individual as a separate cultural layer and differentiate the set from already mentioned ones by adding a biological layer (see table 4).

Table 4. Targowski and Metwalli's Cultural Layers, Influencing the Business Communication Process.

| Cultural layer | Description |
| --- | --- |
| *Global Culture* | Globalization is the driving force here, since it is assumed that people from different cultures rely on the same rules and behaviours to ensure a certain extent of success during communication process in business contexts, i.e. the participants try to adapt to the intercultural situation setting by deliberately deviating from otherwise their dominating cultural behaviours. |
| *National Culture* | It is oriented to traditions, behaviours, feelings, values, etc., that are common to a nation. |
| *Regional Culture* | It embodies the values that individuals share to some extent within a region. |
| *Organizational Culture* | Its meaning is limited to a management means to control organizational performance. |
| *Group Culture* | It refers to a group of people, united by a common relationship as work, profession or family. |
| *Personal Culture* | It represents an individual's specific understanding of time, space and reality. |
| *Biological Culture* | It outlines the universal reactions by humans to their physical needs. |

But Targowski and Metwalli (2003) go even further to describe the general structure of a cultural layer, revealing the intricacy of its contents and relations (see figure 10). The scientists claim that all identified elements in the structure are present at each of the seven levels in spite of the observed inequality in their prominence. The researchers specify "communication channel", i.e. the medium of the message, and "climate", i.e. people's openness to communicate, as other two equally significant factors, concerning the intercultural process.

### 3.7. O'Neil's Cultural Levels

O'Neil (2006) uses simultaneously the terms layers and levels in his framework to propose the existence of three components, forming human learned behavior patterns and perceptions, starting with the most obvious one, as follows:

♦ **The specific body of cultural traditions for a society**. People from separate nationalities may be referred to a shared language, specific traditions and beliefs that differentiate them from other peoples. The majority of members in a nation have acquired their culture from their predecessors.

♦ **The subculture is the second cultural layer, forming human identity.** The complex, diverse societies are viewed as a compound of immigrant groups from different parts of the world whose members often preserve much of their original cultures. The specific shared cultural traits allow a certain group to be identified as a subculture in the new society its members have joined. This classification of cultural levels uses the society of the USA as an example: Vietnamese Americans, African Americans, and Mexican Americans, to describe differences among these subcultures by identity, food tradition, dialect or

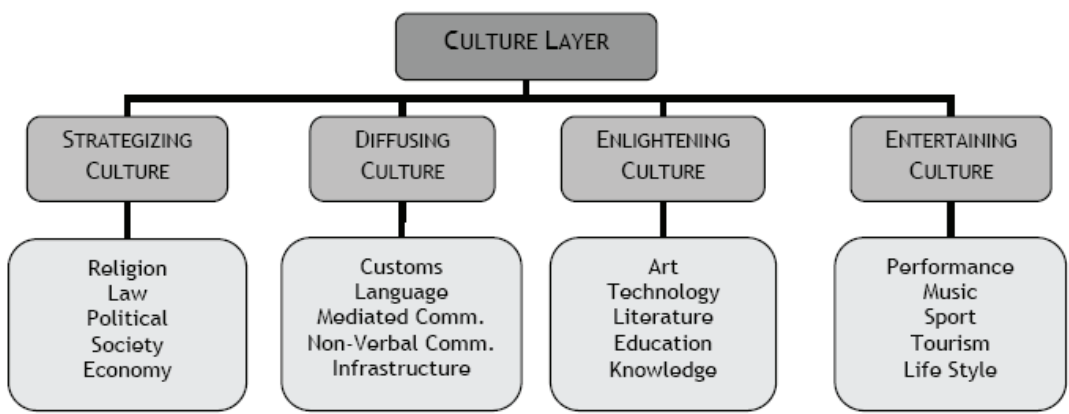

*Fig.10. Structure of a Culture Layer*

*Source: Targowski and Metwalli (2003)*

language, and etc., inherited through common ancestral background and experience. O'Neil dwells on a typical situation in the USA society when the cultural differences between a subculture and the dominant national culture gradually blur and eventually disappear. The scientist outlines the transition of a given subculture into a group of people who identify themselves as citizens of the USA first and claim only a common ancestry (for example German Americans and Irish Americans).

◆ **Cultural universals are the attributes, constituting the third cultural level**. O'Neil defines them as learned behavior patterns, shared by the whole mankind and proposes a long list of such "human cultural" traits, as follows: (a) communicating with a verbal language consisting of a limited set of sounds and grammatical rules for constructing sentences; (b) using age, gender, marriage and descent relationships to classify people (e.g., teenager, senior citizen, woman, man, wife, mother, uncle, cousin); (c) raising children in some sort of family setting; (d) having a concept of privacy; (e) having rules to regulate sexual behavior; etc. The scientist notes that there is a great diversity in the way of carrying out or expressing cultural universals. For instance, people with disabilities (deaf and dumb) use the finger alphabet to communicate with the sign language instead of verbal language. But both types of languages have their specific grammatical rules.

### 3.8. Steven Kaminski's View to Cultural Levels

Kaminski (2006) analyzes cultural levels as pairs, each one consisting of a super-culture and a subculture. A super-culture is defined as "an even more extensive shared perspective that in some way governs the perspective of the subcultures within it" (for example American culture versus General American business culture), while a subculture is defined as "a shared perspective within a larger culture". The analysis here is concentrated on the

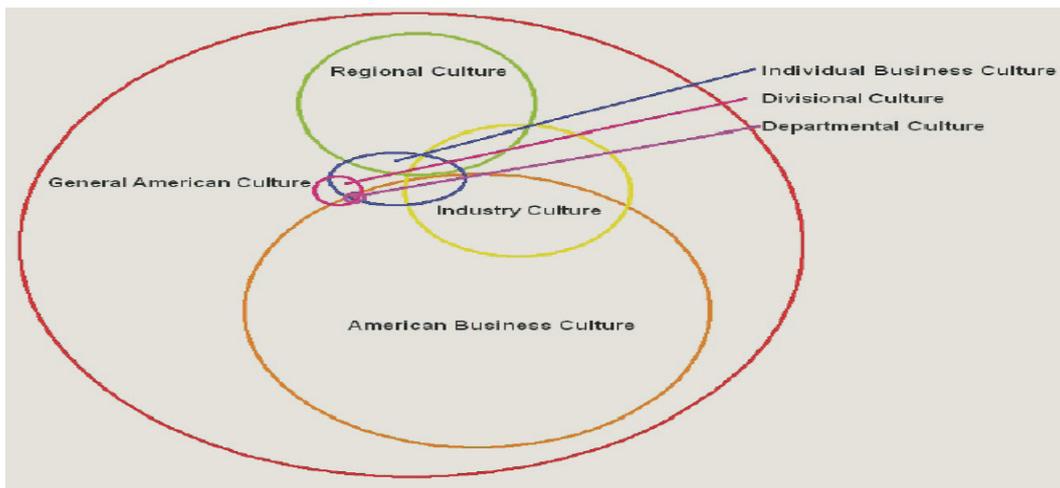

*Fig.11. Kaminski's View to Cultural Levels*                    *Source: (2006)*

*Table 5. Alignment of Cultural Levels by Ali and Brooks.*

| Levels | Descriptions |
|---|---|
| National level (macro level, social level, cross-cultural level) | Culture, shared among the people in a society or a country. |
| Organizational level | Culture, shared among people, working in an organization. |
| Group level | Culture, shared among people with similar profession or occupation, or a subculture of people with specific interests (a political party, a social stratum). |
| Individual level (micro culture, subjective culture) | Subjective culture of the individual – it embodies the extent to which the individual perceives (absorbs, learns) different cultures the last belongs to. |
| Source: Ali and Brooks (2009). | |

relations between the cultural attributes within a chosen pair of adjacent levels. The scientist describes two common characteristics of subcultures, as follows:

♦ Emerging without any influence or direct leadership, exerted by the overarching super culture.

♦ Modifying and/or ignoring key elements of the overarching super culture.

The researcher concludes that better understanding the levels of super-cultures and subcultures may be useful to decision-makers in bringing to surface important assumptions that may generate opportunities for exerting deliberate influence or may be the targets of a change program of a certain business related, social and religious issue (see figure 11).

### 3.9. A Framework of Cultural Levels by Ali and Brooks

Ali and Brooks (2009) make a difference between cultural levels and cultural layers (Hofstede's practices and values). The identified tendency of existing correspondence and relations between levels and layers is similar to Hofstede's ideas of the changing contents in "values – practices" mix. The proposed framework of cultural levels by the scientists is based on a number of cultural studies predominantly in the information technologies sphere where individuals are labeled not only as the basic cultural unit, but also constitute the least general cultural level (see table 5).

*Table 6. Edgar Schein's View on Cultural Levels beyond the Organization.*

| Culture | Category |
|---|---|
| *Macrocultures* | Nations, ethnic and religious groups, occupations that exist globally |
| *Organizational cultures* | Private, public, non-profit, government organizations |
| *Subcultures* | Occupational groups within organizations |
| *Microcultures* | Microsystems within or outside organizations |
| Source:  Schein (2010). | |

*3.10. Edgar Schein's Try in Identification of Cultural Levels beyond the Organization*

Traditionally, Edgar Schein's attention is directed to deeper study and measurement of culture only at organizational level (Schein, 2004; Schein 1999; Schein 1997; Schein to outline his contributions within wider cultural constructs (see table 6). That is why he labels cultural levels beyond the organization as "categories of culture" and analyzes them briefly in the introduction of his book (Schein, 2010). The author describes microsystems as small 1988). But lately the professor feels forced to take into account the increasing importance of cultural levels beyond the organization and without undertaking a new direction of his dominating scientific interest, i.e. the organization, the researcher takes an additional step coherent units within the organizations (surgical teams, task forces), possessing a specific "microculture" whose members cut across occupational groups which in fact differentiates these units from occupational subcultures.

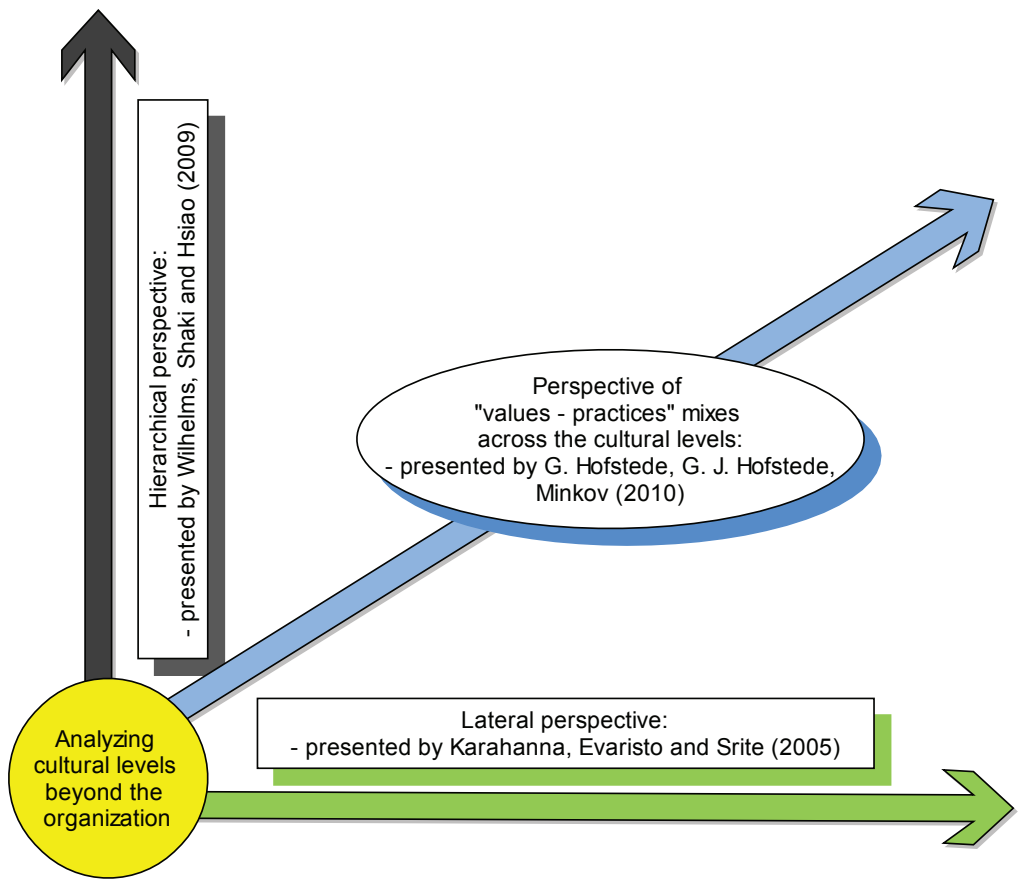

*Fig.12. A Model of Exploring Cultural Levels beyond the Organization.*

## 4. Model of Exploring (Teaching) Cultural Levels beyond the Organization

The drawn canvas of frameworks is rich enough to be considered as complicated and time consuming to be used in daily business, administrative or teaching activities. That is why a new model of exploring and teaching the cultural levels beyond the organizational one is proposed to assist business entities, public sector and non-governmental organizations in their continuous efforts in monitoring their environments and in searching and maintaining long-term, predictable partnerships not only within the EU, but also with "players" from other important regions of the world. Also the model may be useful to students and employees of Bulgarian origin, self-managing their careers, whose cultural awareness is not satisfactory and who inevitably are faced with the motley rag of the European labor market at the latest from the 1st of January 2014. The model consists in elements (concrete cultural levels and layers) that may be analyzed in three perspectives with the potential to provide a detailed snapshot of target environments (driving forces, key players, cultural attributes, etc.) and dynamics of the interaction among separate cultural levels and layers (see figure 12), as follows:

♦ *Hierarchical perspective* (traditional). The analysis is based predominantly on Wilhelms, Shaki and Hsiao' framework (2009) as the most detailed and corresponding to Bulgaria's location at a crossroad between differing nations, religions, continents, etc. The use of this framework permits each observer to fill up its structure with some appropriate contents from supplementary scientific deliverables in the field by Erez and Gati (2004), Hofstede (2010a), Paunov (2005), Trompenaars and Hampden-Turner (2008), Espinar (2010) and O'Neil (2006) according to his/her specific necessities. Additionally, some elements in the structure of a target cultural layer may be identified with the help of Targowski and Metwalli's scientific results (2003) (see figure 10).

♦ *Lateral perspective*. The analysis is based on Karahanna, Evaristo, Srite (2005), but at appropriate cases the frameworks by Rutherford and Kerr (2008) and Kaminski (2006) may be applied, too. This perspective may be exceptionally useful for exploring multinational organization's endeavors in our region, because a number of these, operating in Bulgaria, do not originate from the EU, although some are registered there.

♦ *Perspective of varying "values – practices" mixes* as proposed by Hofstede et. al. (2010b) in order to reveal some aspects of the specific strength of key influencers on separate cultural levels (layers versus levels).

Thus, valuable information, concerning complex existence of target organizations across separate cultural levels may be gathered, appropriately retrieved and wisely used during the timely updating of leadership decision-making and activities in the process of solving the key cultural issues members in the organizations confront everyday.

## 5. Conclusion

Cultural levels beyond the entity proved not to be a prime interest for the researchers at UNWE. But a higher cultural awareness, concerning the cultural levels beyond the organization, is urgently needed by Bulgarian business leaders in their search for new opportunities not only within the Uniform European market, but also on a worldwide scale, especially during the hard times of the World financial and economic crisis. This aim may not be met only with the traditional culture level teaching base, implemented at the University predominantly through Paunov's works (2005, 2008). The new turbulent and more complex business conditions require an elaborated approach of analyzing all already identified and emerging cultural levels, providing:

♦ Needed extent of further segmentation for traditionally considered ones.

♦ Simultaneously three perspectives of analysis, i.e. a hierarchical one, a lateral one and the dynamics of "values-practices" mixes.

♦ Diverse concepts of individual's performance and contribution across cultural levels.

## References

Aleksandrova, M., 2005. Predpriemacheskata orientatsia v konteksta na natsionalnata kulturna sreda. Ikonomicheski alternativi, Br. 3/2005), [online] Available at: http://alternativi.unwe.acad.bg/ [Accessed March, 18th, 2012].

Ali M., Brooks L., 2009. A Situated Cultural Approach for Cross-Cultural Studies in IS. Journal of Enterprise Information Management 22, No. 5/2009, pp. 548-563, Emerald Group Publishing Limited, DOI 10.1108/17410390910993536, www.emeraldinsight.com/1741-0398.htm;

Andreeva, T., The Cultural Industries in the Countries of Southeast Europe and Their Economic Impact in the Context of Social Transformation, Issue 1EN/2008 (Andreeva, Ts., 2008. Kulturnite industrii v stranite ot yugoiztochna Evropa i tyahnoto ikonomichesko vazdeistvie v usloviyata na sotsialna transformatzia. Br. 2/2008), [online] Available at: http://alternativi.unwe.acad.bg/ [Accessed March, 18th, 2012].

Arthur M., Rousseau D., 1996. The Boundariless Career: A New Employment Principle for a New Organizational Era. New York: Oxford University Press.

Chankova, L., 2001. Firmenata kultura – osnova za efektivna inovatsionna deinost. Br. 7-8 (41-42)/2001, 55-57).

Davidkov, Ts., 2005. Kultura na organizatsiite v Bulgaria. Sofia: Universitetsko izdatelstvo "Sveti Kliment Ohridski".

Dimitrov, D., 2005. Konfliktologiya i konfliktologichna kultura. Br.4/2005, [online] Available at: http://alternativi.unwe.acad.bg/ [Accessed March, 18th, 2012].

Dimitrov, D., 2000. Individualizmat na novite eliti u nas i razvitieto na grazhdanskoto obshtestvo. Br.1-3 (25-27)/ 2000, 59-63.

Dimitrov, K., 2009. Several Norms and Beliefs, Defining the Attitude to Human Resources in the Industrial Organizations. Issue 2EN/2009 (Dimitrov, K., 2009. Normi i vyarvania, opredelyashti otnoshenieto kam choveshkite resursi v industrialnite organizatsii. Br.2/2009), [online] Available at: http://alternativi.unwe.acad.bg/ [Accessed March, 18th, 2012].

Dimov, M., 2010. Kade sme nie v evropeiskoto virtualno prostranstvo? Br.4/2010, [online] Available at: http://alternativi.unwe.acad.bg/ [Accessed March, 18th, 2012].

Erez, M., Gati, E., 2004. A Dynamic, Multi-Level Model of Culture: From the Micro Level of the Individual to the Macro Level of a Global Culture. Applied Psychology: an International Review, 2004, *53* (4), 583 –598;

Espinar, A. L., 2010. Intercultural Business Communication:Ttheoretical Framework and Training Methods. AnMal Electrᐗnica 28/2010 Universidad de Cᐗrdoba, (https://gw.uma.es/webmail/src/compose.php?send_to=ff1laesa%40uco.es), ISSN 1697-4239;

Genov, J., 2004. Zashto tolkova malko uspyavame. Sofia: Izdatelstvo "Klasika I stil".

Hofstede, G., 2010a. Levels of Culture. [on-line] Available at: http://www.geerthofstede.nl/culture.aspx,/ [Accessed March, 20th, 2012].

Hofstede, G., Hofstede, G. J., Minkov, M., 2010b. Cultures and Organizations. Software of the Mind. Intercultural Cooperation and Its Importance for Survival, New York: McGraw Hill.

Hofstede, G., 2001. Culture's Consequences. 2nd ed. Newbury Park, CA: Sage Publications,

Hofstede, G. 1994. Cultures and Organisations. London: HarperCollins.

Hofstede, G., 1991. Cultures and Organizations: Software of the Mind, London: McGraw Hill.

Hofstede, G., Neuijen, B., Ohavy, D., Sanders, G., 1990. Measuring Organizational Cultures. Administrative Science Quarterly, 35, pp. 286-316.

Ivanov, P., Durankev, B., Marinov, M., Katrandzhiev H., Stoyanova, M., 2001. Firmenata kultura v Bulgaria. Alternativi, prilozhenie, Sofia.

Kaminski, S. H., 2006. Cultural Strategies of Organizational Design. Bob Jones University, December 23, 2006, [online student resources] Available at: http://www.shkaminski.com/Classes/BJU_MBA_665/LectureNotes/CP%20Chapter%204.htm/ [accessed May, 2nd, 2012].

Karahanna, E., Evaristo, J. and Srite, M., 2005. Levels of Culture and Individual Behaviour: an Integrative Perspective. Journal of Global Information Management, 13, No. 2, April – June, pp. 1-20, ProQuest database.

Kolev, B., Rakadzhiyska, T., 2009. A Tendency toward New Cultural Attitudes of Business Agents in Bulgaria. Issue 2EN/2009 (Kolev, B., Rakadzhiyska, T., 2009. Kam novi kulturalni naglasi na biznes subektite u nas. Br.3/2009), [online]

Available at: http://alternativi.unwe.acad.bg/ [Accessed March, 18th, 2012].

Kolev, B., Rakadjiiska, T., Stoyanova M., Todorova, S., 2009a. Transformation and Adaptation of Bulgarian Business in the Process of Bulgarian Society's Euro-integration. A university project N "SRA" 21.03 – 10/2005 (UNWE).

Lewis R. D., 2006. When Cultures Collide. Leading Across Cultures. Nicholas Brealey International, London.

Milkov, L., 2006. Informatsionnata kultura na prepodavatelya kato factor za sazdavane na kachestvena obrazovatelna usluga vav vissheto uchilishte. Br.1/2006, [online] Available at: http://alternativi.unwe.acad.bg/ [Accessed March, 18th, 2012].

Minkov, M., 2002. Zashto me razlichni? Mezhdukulturni razlichiya v semeystvoto, obshtestvoto i biznesa. Sofia: Izdatelstvo "Klasika I stil".

O'Neil, D., May, 26th, 2006. What is Culture?[online] Available at: http://anthro.palomar.edu/culture/credits.htm/ [Accessed May, 2nd, 2012].

Parusheva, T., 2005. Destinatsia Bulgaria v konteksta na sotsiokulturnite efekti ot evrointegratsiata. Br.1/2005, [online] Available at: http://alternativi.unwe.acad.bg/ [Accessed March, 18th, 2012].

Paunov, M., 2009. Tsennostite na balgarite. Savremenen portret na evropeyski fon. Sofia: Universitetsko izdatelstvo "Stopanstvo".

Paunov, M., 2008. Organizatsionno povedenie i korporativna kultura. Sofia: Universitetsko izdatelstvo "Stopanstvo".

Paunov, M., 2005. Organizatsionna kultura. Sofia: Universitetsko izdatelstvo: "Stopanstvo".

Paunova, M., 2007. Kak da harakterizirame i otsenim kulturata na organizatsiata. Br.6/2007, [online] Available at: http://alternativi.unwe.acad.bg/ [Accessed March, 18th, 2012].

Peycheva, M., 1999. Neobhodimost ot zashtita na etikata na firmata. Br. 23-24/1999, 55-56.

Reeder, K., Macfayden, L. P., Roche, J., & Chase, M., 2004. Negotiating Cultures in Cyberspace: Participation Patterns and Problematics. *Language Learning & Technology,* 8(2), 88-105.

Roth, K., 2012. Chalgiziraneto na Bulgaria – a part of a symposium of German Association for South-East Europe, newspaper Kapital Daily, April, 4th, 2012 (Friday), pp.3, 77.

Rutherford, A. G., Kerr, B., 2008. An Inclusve Approach to Onlne Learnng Envronments: Models and Resources. Turkish Online Journal of Distance Education-TOJDE, April/2008 ISSN 1302-6488, Vol. 9, Number 2, Article 2.

Schein E., 2010. *Organizational culture and leadership*, 4th, ed. JOSSEY – BASS.

Schein E., 2004. *Organizational culture and leadership*, 3rd ed. JOSSEY – BASS.

Schein E., 1999. The Corporate Culture Survival Guide. John Wiley & Sons.

Schein, E., 1997. Organizational Culture and Leadership", 2nd ed. San Francisco: Jossey-Bass;

Schein, E., H., 1988. Organizational Culture and Leadership. San Francisco: Jossey-Bass.

Spasov, T., 2002. Institutsionalnata promyana I ikonomicheskiyat prehod. Br. 5-7 (49-51)/2002, 5-7.

Stavrev, S., 2002. Absurdite na Bulgaria sa normalni, no samo za nas. Br.1-2 (45-46)/2002, 30-31.

Straub, D., Loch, K., Evaristo, R., Karahanna, E. and Srite, M., 2002. Toward a Theory-Based Measurement of Culture. Journal of Global Information Management, Vol. 10 No. 1, pp. 13-23.

Targowski, A., Metwalli, A., 2003. A Framework for Asymmetric Communication Across Cultures. *Dialogue and Universalism*, 7-8, pp. 49-67.

Todorov, K., 2006. Strategii na vodene na pregovori v multikulturna biznes sreda. Br. 5, 2006, [online] Available at: http://alternativi.unwe.acad.bg/ [Accessed March, 18th, 2012].

Todorov, K., 2000. Otnoshenieto strategiya – struktura – organizatzionna kultura v malkite i sredni firmi (MSF). Br.13 (25-27)/2000, 6-13.

Todorov, K., et al., 1992. Firmena kultura I firmeno povedenie. Izdavelstvo "Vek 22".

Trompenaars, F. C. Hampden-Turner, 1998. *Riding the Waves of Culture. Understanding Cultural Diversity in Business*. Nicholas Brealey Publishing London, ISBN 1-85788-176-1.

Tylee, J., 2002. Cultural Issues and the Online Environment. Australian Society for Educational Technology, *International Education and Technology Conference*. Retrieved August 31, 2007, [online] Available at: http://www.csu.edu.au/division/landt/resources/documents/cultural_issues.pdf/ [Accessed May, 2nd, 2012].

Wilhelms, R. W., Shaki, M. K., and Hsiao, C., 2009. How We Communicate about Cultures. A Review of Systems for Classifying Cultures, and a Proposed Model for Standardization. Competitiveness Review: An International Business Journal, Vol. 19 No. 2, 2009, pp. 96-105, Emerald Group Publishing Limited, 1059-5422, DOI 10.1108/10595420910942261.

Yankulov, J., 2006. Izsledvane organizatsionnata kultura na targovskite firmi v Bulgaria. Br. 6, 2006, [online] Available at: http://alternativi.unwe.acad.bg/ [Accessed March, 18th, 2012].

Zlatev, V., 1997. Problemi i predizvikatelstva pred industrialniya menidzhmant v Bulgaria. Br.6/May, 1997, 13-15.